\documentclass[twoside]{article}
\usepackage{fleqn,espcrc2.7,epsf}

\title{Two-colour QCD at finite fundamental quark-number density and related 
       theories.\thanks{Talk presented by D.~K.~Sinclair.}}

\author{S.~J.~Hands,\address[ws]{Department of Physics, University of Wales 
        Swansea, Singleton Park, Swansea SA2 8PP, UK}%
        J.~B.~Kogut,\address{Department of Physics, University of Illinois,
        1110 West Green Street, Urbana, IL 61801-3080, USA}%
        \thanks{Supported in part by the National Science Foundation under
                grant NSF-PHY96-05199.} 
        S.~E.~Morrison,\addressmark[ws]
        and
        D.~K.~Sinclair.\address{HEP Division, Argonne National Laboratory, 9700
        South Cass Avenue, Argonne, IL 60439, USA}
        \thanks{Supported by the U.~S. Department of Energy under grant
                W-31-109-ENG-38.}}

\begin{document}

\begin{abstract}
We are simulating $SU(2)$ Yang-Mills theory with four flavours of dynamical
quarks in the fundamental representation of $SU(2)$ `colour' at finite
chemical potential, $\mu$ for quark number, as a model for QCD at finite baryon
number density. In particular we observe that for $\mu$ large enough this
theory undergoes a phase transition to a state with a diquark condensate which
breaks quark-number symmetry. In this phase we examine the spectrum of light
scalar and pseudoscalar bosons and see evidence for the Goldstone boson
associated with this spontaneous symmetry breaking. This theory is closely
related to QCD at finite chemical potential for isospin, a theory which we are
now studying for $SU(3)$ colour.
\end{abstract}

\maketitle

\section{Introduction}

QCD at finite baryon number density describes nuclear matter including neutron
stars and heavy nuclei. Nuclear matter at finite temperature existed in the
early universe and may be observed in relativistic heavy-ion collisions at CERN
and RHIC.

QCD at finite chemical potential, $\mu$ for quark number has a complex fermion
determinant. Hence current simulation algorithms fail for this theory. Working
explicitly at finite baryon number density trades this complex determinant for
a sign problem.

We therefore turn to studying models which have {\it some} of the properties
of QCD at finite $\mu$. One such property suggested for QCD at finite $\mu$ is
diquark condensation. It is this which led us to study 2-colour ($SU(2)$) QCD
with quarks in the fundamental representation of $SU(2)_{colour}$, which has a
real non-negative determinant (and pfaffian), permitting simulations.

2-colour QCD with fundamental quarks is confining and has a sensible meson
spectrum but an unphysical ``baryon'' spectrum. At $m=0$ and $\mu=0$ the
usual $SU(N_f) \times SU(N_f) \times U(1)$ chiral symmetry is enlarged to 
$SU(2 N_f)$. The ``meson'' multiplets are enlarged to include 2-quark 
``baryons''. Spontaneous breakdown of chiral symmetry gives 
$SU(2 N_f) \rightarrow Sp(2 N_f)$ rather than the usual 
$SU_R(N_f) \times SU_L(N_f) \times U_V(1)\rightarrow U_V(N_f)$. For large
enough chemical potential we expect a spontaneous breakdown of quark number
with a diquark condensate. This condensate is a colour singlet and has
associated Goldstone bosons \cite{qcd2}. For true QCD the condensate is, of
necessity, coloured, and breaks the gauge symmetry in the Higgs manner --
colour superconductivity. For a chiral perturbation theory analysis of
2-colour QCD at finite $\mu$ see \cite{chiral}.

The 2 flavour version of this theory can be reinterpreted as 2-colour QCD at
finite chemical potential for isospin. However, it is easy to see that we can
treat true (3-colour) QCD at finite chemical potential for isospin since it
has a real, non-negative determinant. Since nuclear matter has finite (negative)
isospin density as well as finite baryon number density, this theory describes
a surface in the phase diagram for nuclear matter. For a discussion of this
theory including a chiral perturbation theory analysis see \cite{stephanov}.

\section{Lattice 2-colour QCD at finite quark-number chemical potential}

The standard staggered fermion transcription of this theory is:
\begin{equation}
S_f = \sum_{sites}\left\{\bar{\chi}[D\!\!\!\!/\,(\mu) + m]\chi 
+ \frac{1}{2}\lambda[\chi^T\tau_2\chi + \bar{\chi}\tau_2\bar{\chi}^T]\right\}
\end{equation}
where the chemical potential $\mu$ is introduced by multiplying links in the
$+t$ direction by $e^\mu$ and those in the $-t$ direction by $e^{-\mu}$. The
diquark source term (Majorana mass term) is added to allow us to observe
spontaneous breakdown of quark-number on a finite lattice.

Integrating out the fermion fields gives:
\begin{equation}
pfaffian\left[\begin{array}{cc} \lambda\tau_2     &    {\cal A}       \\
                                     -{\cal A}^T        &    \lambda\tau_2
\end{array}\right] = \sqrt{{\rm det}({\cal A}^\dagger {\cal A} + \lambda^2)}
\end{equation}
where
\begin{equation}
           {\cal A} \equiv  D\!\!\!\!/\,(\mu)+m
\end{equation}
Note that the pfaffian is strictly positive, so that we can use the hybrid 
molecular dynamics method to simulate this theory using ``noisy'' fermions to
take the square root, giving $N_f=4$.

When $m = \lambda = \mu = 0$, the chiral symmetry of the above action is 
expanded to $U(2)$, from the $U(1) \times U(1)_V$ of true lattice QCD. 
This breaks spontaneously -- $U(2) \rightarrow U(1)_V$ -- yielding 3 Goldstone
bosons. For $m = \lambda = 0$, $\mu \ne 0$ we expect spontaneous breakdown of
quark number and 2 Goldstone bosons -- a scalar diquark and a pseudoscalar 
diquark. For $\lambda = 0$, $m \ne 0$, $\mu \ne 0$ we have no spontaneous
symmetry breaking for small $\mu$. For $\mu$ large enough ($\mu > m_\pi/2$ ?)
we expect spontaneous breakdown of quark number and one Goldstone boson -- a
scalar diquark. (See \cite{qcd2}  for a full discussion of these symmetries,
and for early simulations of the 8 flavour theory at $\lambda=0$.)

We are simulating this $N_f=4$ theory on an $8^4$ lattice, measuring the chiral
and diquark condensates, the fermion number density, the Wilson/Polyakov line,
etc.. We are repeating these simulations on $12^3 \times 24$ lattices, where,
in addition, we are measuring all local scalar and pseudoscalar meson and 
diquark propagators (connected and disconnected).

\section{Results}

We have preliminary results for a relatively large quark mass, $m=0.1$,
and $\lambda=0.01,0.02$ (and $0$ for small $\mu$). We have recently started
simulations at a small quark mass, $m=0.025$ and $\lambda=0.0025,0.005$ where we
should observe a more complex spectrum of Goldstone and pseudo-Goldstone
bosons. Finally we will simulate at $m=0$ where there should be 3 Goldstone
bosons for $\mu=0$ and 2 Goldstone bosons for $\mu > 0$. All our simulations
are done at $\beta=1.5 \approx \beta_c(N_t=4)$.

\begin{figure}[htb]
\epsfxsize=3in
\centerline{\epsffile{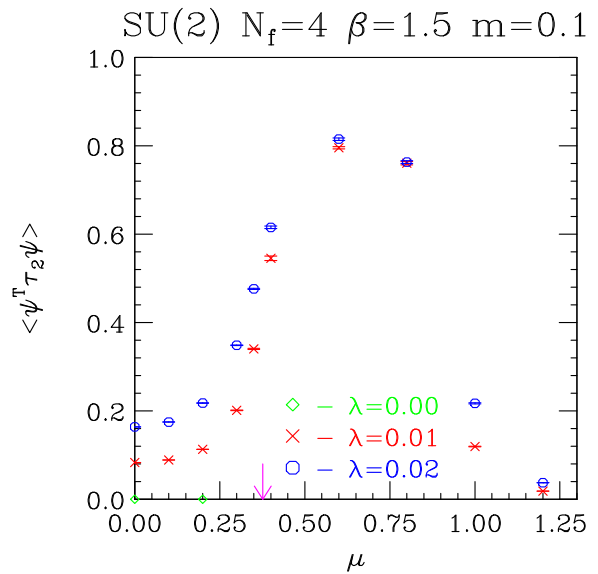}}
\caption{Diquark condensate as a function of $\mu$ on an $8^4$ lattice. 
The arrow indicates $\mu \approx m_\pi/2$.}
\label{fig:pt2p}
\end{figure}

Figure \ref{fig:pt2p} shows the diquark condensate 
$\langle\chi^T\tau_2\chi\rangle$ as a function of chemical potential
potential.
For small $\mu$ there is no diquark condensate and quark-number is a good
symmetry. At $\mu = \mu_c \sim m_\pi/2$ there is a phase transition, above
which there is a diquark condensate, and quark-number is spontaneously broken.
This condensate increases to a maximum and then falls towards zero at large
$\mu$.

\begin{figure}[htb]
\epsfxsize=3in                                                                
\centerline{\epsffile{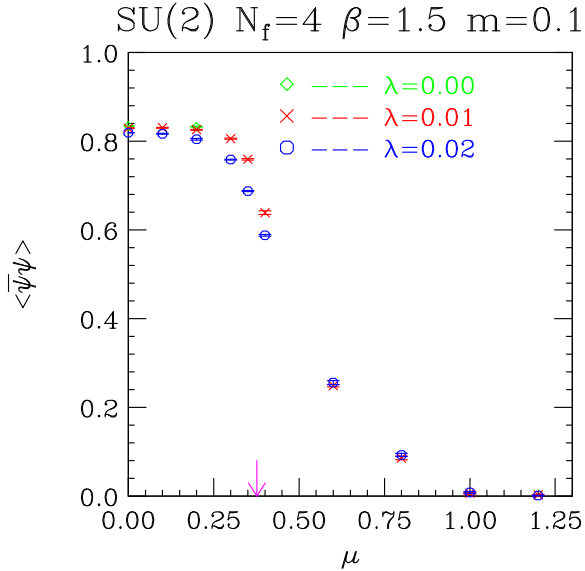}}                                              
\caption{Chiral condensate as a function of $\mu$ on an $8^4$ lattice.}
\label{fig:pbp}
\end{figure}

Figure \ref{fig:pbp} shows the chiral condensate, $\langle\bar{\chi}\chi\rangle$
as a function of $\mu$. 
The chiral condensate is approximately constant for $\mu < \mu_c$. For $\mu >
\mu_c$ it falls, approaching zero for large $\mu$.

\begin{figure}[htb]
\epsfxsize=3in     
\centerline{\epsffile{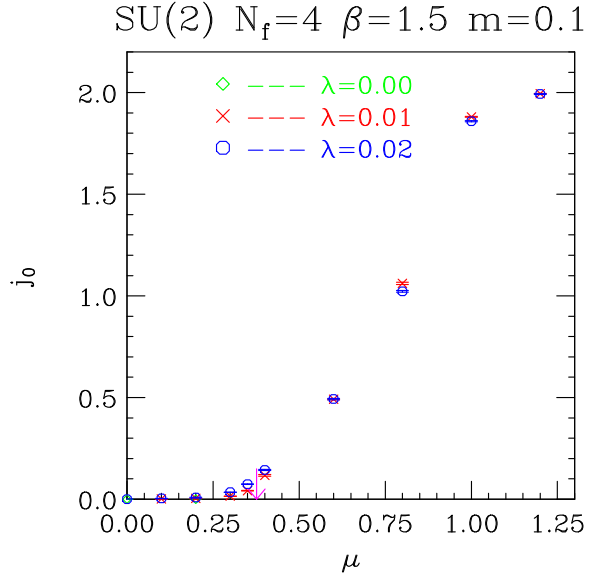}}
\caption{Quark-number density as a function of $\mu$ on an $8^4$ lattice.}
\label{fig:j0}
\end{figure}

Figure \ref{fig:j0} shows the quark-number density as a function of $\mu$.
The quark-number density $j_0$ is zero for $\mu < \mu_c$ and increases for
$\mu > \mu_c$ approaching the saturation value $j_0=2$ for large $\mu$.

\begin{figure}[htb]                 
\epsfxsize=3in     
\centerline{\epsffile{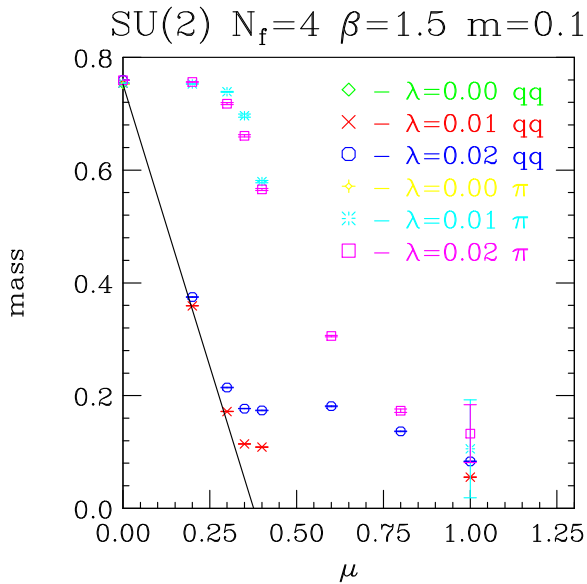}}
\caption{Pion ($\pi$) and scalar diquark ($qq$) masses as functions of $\mu$.
The straight line is $mass=m_\pi - 2\mu$. The arrow is at $\mu=m_\pi/2$.}
\label{fig:mass}
\end{figure}

Finally, in figure~\ref{fig:mass} we show the masses of the pion and the
scalar diquark state which is orthogonal to the diquark condensate, as functions
of $\mu$. This is from the $12^3 \times 24$ lattice where we have less complete
`data'.
The mass of the scalar diquark falls roughly as the expected $m_\pi - 2\mu$
as $\mu$ is increased towards $\mu_c$. For $\mu > \mu_c$ it is small enough
and its $\lambda$ dependence is such that one can believe that it would become
massless as $\lambda \rightarrow 0$. Thus the scalar diquark does appear to be
the Goldstone boson associated with spontaneous quark-number breaking. The
pion mass appears to remain constant for $\mu < \mu_c$, dropping towards zero
as $\mu$ is increased beyond $\mu_c$. This mass spectrum and the behaviour
of the order parameters is in good agreement with the predictions from chiral
perturbation theory for $\mu \le 0.6$.

\section{\bf QCD at finite isospin density}

For QCD at finite chemical potential for isospin ($I_3$), the staggered quark
action is
\begin{equation}
S_f = \sum_{sites}\left\{\bar{\chi}[D\!\!\!\!/\,(\tau_3\mu) + m]\chi
                        + \lambda\bar{\chi}i\tau_2\epsilon\chi\right\}
\end{equation}
where the explicit symmetry breaking term proportional to $\lambda$ is needed
to observe spontaneous isospin breaking on finite lattices; $\tau_i$ are isospin
matrices acting on the isodoublet $\chi$, and $\epsilon=(-1)^{x+y+z+t}$. 
Integrating out the fermion fields yields:
\begin{equation}
           {\rm det}({\cal A}^\dagger {\cal A} + \lambda^2)            
\end{equation}
where
\begin{equation}
           {\cal A} \equiv  D\!\!\!\!/\,(\mu)+m                                 
\end{equation}                                                        
Note ${\cal A}$ is only a $1 \times 1$ matrix in isospin space. This describes
8 fermion flavours, so we use hybrid molecular dynamics with ``noisy'' fermions
to perform the simulations, to reduce this to 2 flavours. Note this determinant
has exactly the same form as that for 2-colour QCD at finite quark number 
density, except now we can use $SU(3)$ colour. To get a sensible flavour 
interpretation for this 2-flavour theory in the continuum limit requires
making the field redefinition $\chi_2 \rightarrow \xi_5 \chi_2$, where $\xi_5$ 
is the analogue of $\gamma_5$ in $SU(4)$ flavour space. 

\section{Conclusions}

2-colour QCD at finite chemical potential for quark number undergoes a phase
transition at $\mu=\mu_c \le m_\pi/2$. The high $\mu$ phase is characterized by
spontaneous breakdown of quark number precipitated by a diquark condensate,
with a single Goldstone boson. Is there a second transition at high $\mu$ to a
free-field phase? What can we learn from this model about diquark condensates
in true ($SU(3)$) QCD?

We need to simulate at smaller quark masses (including zero) where the
competition between chiral and quark-number symmetry breaking should lead to a
more complex spectrum of Goldstone and pseudo-Goldstone bosons. A more extensive
analysis of the spectrum of scalar and pseudoscalar mesons and diquarks is
called for.

QCD at finite chemical potential, $\mu_I$, for isospin maps a surface in the
phase diagram of nuclear matter which, in addition to having a finite baryon
number density, also has a finite (negative) isospin density. A
reinterpretation of 2-colour QCD at finite quark number density yields
2-colour QCD at finite isospin density, and gives us a guide as to what to
expect in the 3-colour case. At finite $\mu_I$, true (3-colour) QCD has a real
positive determinant which allows us to simulate it. For large enough chemical
potential we should get spontaneous breaking of the remaining $U(1)$ isospin
symmetry generated by $I_3$ with a charged pion condensate and one true
Goldstone pion.

Recently, the work of \cite{qcd2} has been extended to finite temperature as
well as $\mu$ by the Hiroshima group\cite{t+mu}. In addition the eigenvalues of
the Dirac matrix in \cite{qcd2} have been studied by \cite{markum}.
Finally we would like to
mention related work on gauge theories with adjoint fermions \cite{adjoint}.

\section*{Acknowledgements}

These simulations are being performed on the Cray SV1's ($8^4$, $m=0.1$) and
the IBM SP ($12^3 \times 24$) at NERSC, and on the IBM SP's ($8^4$, $m=0.025$)
at NPACI.

\end{document}